\documentclass[preprint,proceedings]{rmaa}

\usepackage{graphicx}

\SetYear{2004}
\SetConfTitle{Compact Binaries in the Galaxy and Beyond}

\title{ Simultaneous multi-wavelength observations of microquasars (the MINE collaboration)} 

\author{
  Y. Fuchs,\altaffilmark{1} 
  J. Rodriguez,\altaffilmark{1,2}
  I.~F. Mirabel,\altaffilmark{1}
  S.~E. Shaw,\altaffilmark{3,2}
  P. Kretschmar,\altaffilmark{4,2}
  M. Rib\'o,\altaffilmark{1}
  S. Chaty,\altaffilmark{5,1}
  V.~Dhawan,\altaffilmark{6}
  I.~Brown,\altaffilmark{7}
  R. Spencer,\altaffilmark{7}
  G.~G. Pooley,\altaffilmark{8}
  D.~C. Hannikainen\altaffilmark{9}}

\altaffiltext{1}{Service d'Astrophysique, CEA/Saclay, France.}
\altaffiltext{2}{{\it INTEGRAL} Science Data Center, Versoix, Switzerland.}
\altaffiltext{3}{University of Southampton, Southampton, UK.}
\altaffiltext{4}{MPE, Garching, Germany.}
\altaffiltext{5}{Universit\'e Paris 7, Paris, France}
\altaffiltext{6}{National Radio Astronomy Observatory, Socorro, USA.}
\altaffiltext{7}{Jodrell Bank, Manchester Univ., UK.}
\altaffiltext{8}{Mullard Radio Astronomy Obs., Cambridge, UK.}
\altaffiltext{9}{University of Helsinki, Helsinki, Finland.}

\shortauthor{Fuchs et al.\@}
\shorttitle{MINE observations of microquasars}

\suppressfulladdresses

\listofauthors{Y. Fuchs, J. Rodriguez, I. F. Mirabel, S.~E. Shaw, 
P. Kretschmar, M. Rib\'o, S. Chaty, V. Dhawan, I. Brown, R. Spencer, G.~G. Pooley \& D.~C. Hannikainen}
\indexauthor{Fuchs, Y.}
\indexauthor{Rodriguez, J.}
\indexauthor{Mirabel, I.~F.}
\indexauthor{Shaw, S.~E.}
\indexauthor{Kretschmar, P.}
\indexauthor{Rib\'o, M.}
\indexauthor{Chaty, S.}
\indexauthor{Dhawan, V.}
\indexauthor{Brown, I.}
\indexauthor{Spencer, R.}
\indexauthor{Pooley, G.~G.}
\indexauthor{Hannikainen, D.~C.}

\abstract{We present the
international collaboration MINE (Multi-$\lambda$ {\it INTEGRAL} NEtwork) aimed
at conducting multi-wavelength observations of 
microquasars simultaneously with the {\it INTEGRAL} satellite. 
The first results on GRS\,1915+105 are encouraging and those to
come should help us to understand the physics of the accretion and
ejection phenomena around a compact object.}

\addkeyword{stars: individual (GRS\,1915+105)}
\addkeyword{X-rays: binaries}

\begin{document}
\maketitle

\section{Introduction}
	Microquasars are X-ray binaries producing relativistic jets
	and thus appear as miniature replicas of distant quasars
	(Mirabel \& Rodr\'\i guez 1999).
	Their emission
	spectra, variable with time, range from the radio to the
	$\gamma$-ray wavelengths.
	We present here the first multi-wavelength campaign on
	GRS\,1915+105 involving the {\it INTEGRAL} satellite (3\,keV--10\,MeV).
	This campaign was conducted by the MINE
	(\mbox{Multi-$\lambda$} {\it INTEGRAL} NEtwork, see
	{\sf http://elbereth.obspm.fr/$\sim$fuchs/mine.html}) international
	collaboration aimed at performing multi-wavelength
	observations of galactic X-ray binaries simultaneously with
	{\it INTEGRAL}.

\section{GRS\,1915+105}
	The microquasar GRS\,1915+105 is extremely variable at all
	wavelengths (see Fuchs et al.\@ 2003a for a review).  It hosts the
	most massive known stellar mass black hole of our Galaxy with
	M=14$\pm$4M$_{\odot}$ (Greiner et al.\@ 2001).
	In the radio 
	it can show superluminal ejections at arcsec scales
	(Mirabel \& Rodr\'\i guez 1994)
	leading to a maximum distance of 11.2$\pm$0.8\,kpc (Fender et al.\@ 1999),
	and  a
	compact jet at milli-arcsecond scales (Dhawan et al.\@ 2000).
	 
\begin{figure}[h]
   \centering
   \includegraphics[angle=-90,width=\columnwidth]{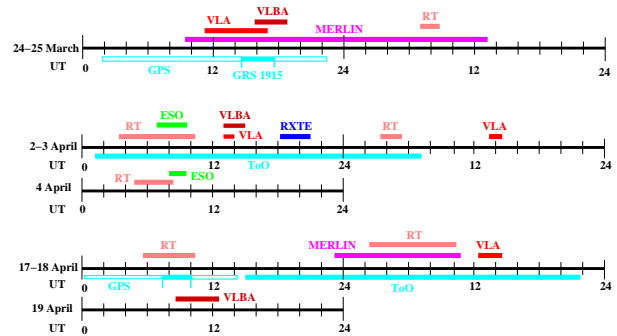}
      \caption{Viewgraph of the observing campaign in spring
      2003, indicating the dates, time and involved observatories (GPS
      = Galactic Plane Survey of {\it INTEGRAL}).}
       \label{figcamp}
\end{figure}

	We conducted a multi-wavelength observation campaign of
	GRS\,1915+105 in spring 2003 (see Figure~\ref{figcamp}).
	Here we focus only on the April observations when
	ToO (Targets of
	Opportunity) were triggered by the MINE collaboration.
	This (nearly) simultaneous campaign
	involved the VLA, the 
	VLBA, MERLIN and the Ryle Telescope (RT) in radio, the ESO/NTT
	in IR, {\it RXTE} and {\it INTEGRAL}
	in X/$\gamma$-rays. More details and description of the 
	April 2 observations can be found in Fuchs et al.\@ (2003b).

\begin{figure}[ht]
   \centering
   \includegraphics[width=6.5cm]{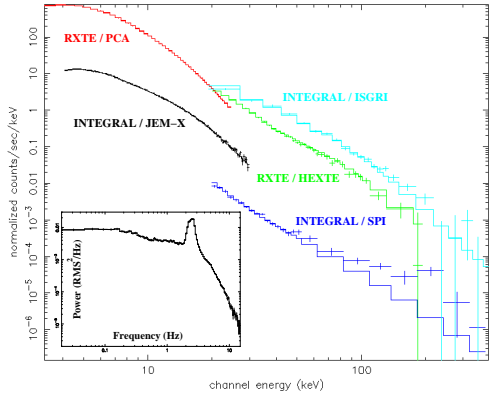}
   \vspace*{-0.3cm}
      \caption{X/$\gamma$-ray spectra and fit of
   GRS\,1915+105 measured with {\it RXTE} \& {\it INTEGRAL}
    on April 2, 2003. 
   The structures at $E$$>$50\,keV in
   the SPI spectrum are instrumental background lines not adequately corrected.
   The PCA power density spectrum (inset)
   shows a clear QPO at 2.5\,Hz.}
       \label{figspec}
\end{figure}

\begin{figure}[ht]
   \centering
   \includegraphics[width=7.0cm]{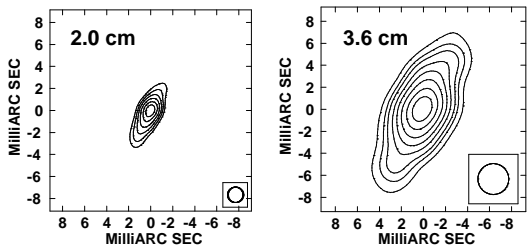}
   \vspace*{-0.3cm}
      \caption{VLBA images at 2.0 \& 3.6\,cm on April 2, 2003 showing
      the compact jet. 
	1\,mas = 12\,AU at 12\,kpc.} 
       \label{figjet}
\end{figure}

	Our
	 observations took place during the \emph{plateau}
	state of GRS\,1915+105 (Fender et al.\@\,1999)
	i.e.\@ quasi-steady
	 {\it RXTE}/ASM (2--12\,keV) flux
	$\sim$50\,cts/s and  high steady radio level 
	($>$100\,mJy).
	On April\,2 (Figure~\ref{figspec}) \& 18 
	the high energy emission 
	shows a power law dominated spectrum ($>$60\% at
	3--20\,keV) with a photon index $\Gamma$=2.9 and 2.75, respectively. 
	This state is much softer 
	than the classical low/hard state of the other BH binaries 
	and is closer to the very high or intermediate states
	(McClintock \& Remillard 2004).
	The {\it INTEGRAL} observations
	show that this power law spectrum extends up to 400\,keV
	without any cutoff during this \emph{plateau} state, consistent with
	the observations with {\it CGRO}/OSSE 
	(Zdziarski et al.\@ 2001).
%
	The estimated luminosity on April 2 \& 18 is respectively
	$\sim$\,$7.5$\,$\times$\,$10^{38}$\,erg\,s$^{-1}$ overestimated using the PCA 
	and $\sim$\,$2.8$\,$\times$\,$10^{38}$\,erg\,s$^{-1}$ using JEM-X, corresponding to
	40\% and 16\% of the Eddington luminosity for a $14M_{\odot}$
	black hole. 
	As shown in Figure~\ref{figspec}, a very
	clear Quasi-Periodic Oscillation (QPO) at 2.5\,Hz with a 14\%
	rms level was observed in the {\it RXTE}/PCA signal, 
	which is
	consistent with the previous observations of the \emph{plateau} state
	(see also Rodriguez et al.\@ 2003).

	The VLBA high resolution images on April 2 (Figure~\ref{figjet}) and 18 show the
	presence of a compact radio jet with a $\sim$7--14\,mas length
	(85--170\,AU at 12\,kpc). 
	The optically thick synchrotron emission from this jet is
	responsible for the high radio levels.
	On April 18 the MERLIN image shows a radio extension ($\sim$0.3$''$) which is 
	probably the trace of a superluminal ejection which occurred
	on April~4.

	On April\,2 the source was fairly bright in near-IR 
	with an excess of 75--85\%
	in the $K_{\mathrm s}$-band compared to the $K$=14.5--15\,mag.\@
	of the K-M giant donor star of the binary.
	According to the spectral energy distribution, this IR excess is 
	compatible with a strong contribution from
	the synchrotron emission of the jet extending from the radio
	up to the near-IR. 
	Different components, however, contribute to the IR
	in addition to the jet,
	such as the donor-star, the external part of the
	accretion disc or a free-free emission.

\section{Conclusions and Prospects}
	Here for the first time, we observed simultaneously all the
	properties of the \emph{plateau} state of GRS\,1915+105\,: 
	a powerful compact
	radio jet, responsible for the strong steady radio emission and
	probably for a significant part of the bright near-IR
	emission, as well as a QPO (2.5\,Hz) in the X-rays 
	and a power law dominated X-ray spectrum with
	a $\Gamma$$\sim$3 photon index up to at least
	400\,keV. Forthcoming works will study
	detailed fits of the X-ray spectra, 
	to determine for example
	whether this power law is due to an inverse Compton
	scattering of soft disc photons 
	on the base of the compact jet 
	or not.
	In order to better
	understand the unusual behaviour of GRS\,1915+105, we need to
	carry out similar simultaneous broad-band campaigns during the
	other states, in particular during the sudden changes 
	 that correspond to powerful relativistic
	ejection events.


\end{document}